\documentclass[10pt,a4paper,english,showpacs,pra,superscriptaddress]{revtex4}

\usepackage{times}
\usepackage{amsmath}
\usepackage{amssymb}
\usepackage[dvips]{graphicx}
\usepackage[unicode=true,pdfusetitle,
 bookmarks=true,bookmarksnumbered=false,bookmarksopen=false,
 breaklinks=false,pdfborder={0 0 1},backref=false,colorlinks=false]
 {hyperref}

\begin{document}

\title{Bohmian trajectories for bipartite entangled states}
\author{A. R de Almeida}
\affiliation{Instituto de F\'isica, Universidade Federal de Goi\'as, 74001-970, Goi\^ania
(GO), Brazil}
\affiliation{UnUCET - Universidade Estadual de Goi\'as, 75132-903, An\'apolis (GO),
Brazil}

\author{M. A. de Ponte}
\affiliation{Universidade Regional do Cariri, Centro de Ci\^encias e Tecnologia,
Departamento de F\'isica, 63010-970, Juazeiro do Norte (CE), Brazil}

\author{W. B. Cardoso}
\affiliation{Instituto de F\'isica, Universidade Federal de Goi\'as, 74001-970, Goi\^ania
(GO), Brazil}

\author{A. T. Avelar}
\affiliation{Instituto de F\'isica, Universidade Federal de Goi\'as, 74001-970, Goi\^ania
(GO), Brazil}

\author{M. H. Y. Moussa}
\affiliation{Instituto de F\'isica de S\~ao Carlos, Universidade de S\~ao Paulo, Caixa
Postal 369, 13560-970, S\~ao Carlos (SP), Brazil}

\author{N. G. de Almeida}
\affiliation{Instituto de F\'isica, Universidade Federal de Goi\'as, 74001-970, Goi\^ania
(GO), Brazil}

\begin{abstract}
We derive Bohm's trajectories from Bell's beables for arbitrary bipartite
systems composed by dissipative noninteracting harmonic oscillators
at finite temperature. As an application of our result, we calculate
the Bohmian trajectories of particles described by a generalized Werner
state, comparing the trajectories when the sate is either separable
or entangled. We show that qualitative differences appear in the trajectories
for entangled states as compared with those for separable states.
\end{abstract}


\maketitle

\section{Introduction}

Entanglement phenomenon is possibly the most striking feature of Quantum
Mechanics, playing a key role in quantum information processing and
quantum computing. The striking feature of entanglement was experimentally
turned possible after the seminal paper by Bell \cite{Bell}, and
since then efforts from experimental and theoretical research to demonstrate
the violation of Bell's inequalities have been undertaken. To unequivocally
demonstrate violation of Bell's inequality, a major development in
experimental techniques has been carried out to produce entangled
photons from the cascade atomic process \cite{Aspect} to the parametric
down-conversion processes \cite{PDC}. Massive entangled particles
have also been produced through radiation--matter interaction in cavity
QED \cite{Haroche} and trapped ions \cite{TI}. In the latter case,
a controlled entanglement of 14 quantum bits has been recently generated
enabling the implementation of the largest quantum register to date
\cite{Insbruck}. The violation of a form of Bell's inequality has
also been verified with massive entangled particles within an ion
trap \cite{VBI}. Parallel to the experimental achievements, theoretical
physics has been struggling with recently advanced striking features
of entanglement, such as the derivation of separability criterion
for density matrices \cite{PH}, entanglement sudden death \cite{Eberly},
and quantum discord \cite{Zurek}.

Bohmian mechanics is a theory equivalent with orthodox quantum mechanics
having the advantage of providing ontological meaning for the quantum
particle trajectory \cite{Holland}. Bohmian mechanics assumes that
the complete description of particle systems is provided by its wave
function $\Psi$ and its configuration $Q=(Q_{1},..,Q_{N})$
$\in\mathbb{R}^{3N}$ , where $Q_{\alpha}$ is the position of
the $\alpha$-th particle. While the wave function $\Psi(Q)$
evolves according to the Schrödinger's equation, the
motion of the particles evolves according to the equation $m_{\alpha}dQ_{\alpha}/dt=\hbar{\rm Im}\left\{ \Psi^{-1}\partial\Psi/\partial Q_{\alpha}\right\} $.
Since only the average trajectories are experimentally accessible,
the particle positions are the \textquotedblleft{}hidden variables\textquotedblright{}
of Bohmian mechanics.

In the present work we focus on entangled states of bipartite dissipative
systems from the Bohmian trajectories perspective as formulated by
Vink's extension of Bell's beables \cite{Bell1,Vink}. Our goal is
to verify what happens to the Bohmian trajectories of a given bipartite
state at the instant occurring separability. To achieve our goal,
we first derive Bohm's trajectories from Bell's beables for arbitrary
bipartite states under thermal reservoirs at finite temperature. On
this regard, we note that an extension of Bell's beables that encompasses
dissipation and decoherence for one particle state has been advanced
\cite{Felipe}, where the diffusive terms in Nelson's stochastic formalism
are naturally incorporated into Bohm's causal dynamics. Summarizing,
here we generalize the approach of Ref.\cite{Felipe} to include dissipative
two-particle states, thus allowing us to study Bohmian trajectories
for correlated quantum systems described either by pure or mixed states
under dissipation at finite temperature.

\section{Bohm's trajectories for two-particles density matrices}

In this section we derive Bohm's trajectories from Bell's beables
to encompass two entangled noninteracting particles under independent
reservoirs at finite temperature, whose master equation is
\begin{eqnarray}
\frac{\partial\rho\left(t\right)}{\partial t} & = & -\frac{i}{\hbar}\left[H,\rho\left(t\right)\right]+\sum_{\alpha}\frac{\gamma_{\alpha}}{2}\Big[(\bar{n_{\alpha}}+1)(2a_{\alpha}\rho a_{\alpha}^{\dagger}-a_{\alpha}^{\dagger}a_{\alpha}\rho-\rho a_{\alpha}^{\dagger}a_{\alpha})\nonumber \\
 & + & \bar{n_{\alpha}}(2a_{\alpha}^{\dagger}\rho a_{\alpha}-a_{\alpha}a_{\alpha}^{\dagger}\rho-\rho a_{\alpha}a_{\alpha}^{\dagger})\Big],\label{ME}
\end{eqnarray}
with $a_{\alpha}$ and $a_{\alpha}^{\dagger}$ being the usual annihilation
and creation operators in Fock spaces, $\gamma_{\alpha}$ is the corresponding
dissipative rate with reservoir average thermal photon number $\bar{n_{\alpha}}$,
and $\alpha=\{1,2\}$ refers to each particle of the system and its
corresponding reservoir. This master equation generates the probability
density $P_{n_{1}n_{2}}\left(t\right)=\left\langle \varphi_{n_{1}},\chi_{n_{2}}\right\vert \rho\left(t\right)\left\vert \varphi_{n_{1}},\chi_{n_{2}}\right\rangle $
such that
\begin{equation}
\hbar\frac{\partial P_{n_{1}n_{2}}}{\partial t}=\left\langle \varphi_{n_{1}},\chi_{n_{2}}\right\vert \frac{\partial\rho\left(t\right)}{\partial t}\left\vert \varphi_{n_{1}},\chi_{n_{2}}\right\rangle \equiv\sum_{m_{1},m_{2}}J_{n_{1}n_{2}m_{1}m_{2}}\text{.}\label{continuity}
\end{equation}
Now, using the Eq. (\ref{ME}) and completeness relations we obtain
\begin{small}
\begin{eqnarray}
J_{n_{1}n_{2}m_{1}m_{2}} & = & 2{\rm Im\Big\{}\left\langle \varphi_{n_{1}},\chi_{n_{2}}\right\vert H\left\vert \varphi_{m_{1}},\chi_{m_{2}}\right\rangle \left\langle \varphi_{m_{1}},\chi_{m_{2}}\right\vert \rho(t)\left\vert \varphi_{n_{1}},\chi_{n_{2}}\right\rangle \Big\}\nonumber \\
 & + & \frac{\hbar}{2}\sum_{k_{1},k_{2},\alpha}\gamma_{\alpha}(\bar{n_{\alpha}}+1)\Big\{2\left\langle \varphi_{n_{1}},\chi_{n_{2}}\right\vert a_{\alpha}\left\vert \varphi_{m_{1}},\chi_{m_{2}}\right\rangle \left\langle \varphi_{m_{1}},\chi_{m_{2}}\right\vert \rho(t)\left\vert \varphi_{k_{1}},\chi_{k_{2}}\right\rangle \left\langle \varphi_{k_{1}},\chi_{k_{2}}\right\vert a_{\alpha}^{\dagger}\left\vert \varphi_{n_{1}},\chi_{n_{2}}\right\rangle \nonumber \\
 & - & \left\langle \varphi_{n_{1}},\chi_{n_{2}}\right\vert a_{\alpha}^{\dagger}\left\vert \varphi_{m_{1}},\chi_{m_{2}}\right\rangle \left\langle \varphi_{m_{1}},\chi_{m_{2}}\right\vert a_{\alpha}\left\vert \varphi_{k_{1}},\chi_{k_{2}}\right\rangle \left\langle \varphi_{k_{1}},\chi_{k_{2}}\right\vert \rho(t)\left\vert \varphi_{n_{1}},\chi_{n_{2}}\right\rangle \nonumber \\
 & - & \left\langle \varphi_{n_{1}},\chi_{n_{2}}\right\vert \rho(t)\left\vert \varphi_{m_{1}},\chi_{m_{2}}\right\rangle \left\langle \varphi_{m_{1}},\chi_{m_{2}}\right\vert a_{\alpha}^{\dagger}\left\vert \varphi_{k_{1}},\chi_{k_{2}}\right\rangle \left\langle \varphi_{k_{1}},\chi_{k_{2}}\right\vert a_{\alpha}\left\vert \varphi_{n_{1}},\chi_{n_{2}}\right\rangle \Big\}\nonumber \\
 & + & \frac{\hbar}{2}\sum_{k_{1},k_{2},\alpha}\gamma_{\alpha}\bar{n}_{\alpha}\Big\{2\left\langle \varphi_{n_{1}},\chi_{n_{2}}\right\vert a_{\alpha}^{\dagger}\left\vert \varphi_{m_{1}},\chi_{m_{2}}\right\rangle \left\langle \varphi_{m_{1}},\chi_{m_{2}}\right\vert \rho(t)\left\vert \varphi_{k_{1}},\chi_{k_{2}}\right\rangle \left\langle \varphi_{k_{1}},\chi_{k_{2}}\right\vert a_{\alpha}\left\vert \varphi_{n_{1}},\chi_{n_{2}}\right\rangle .\nonumber \\
 & - & \left\langle \varphi_{n_{1}},\chi_{n_{2}}\right\vert a_{\alpha}\left\vert \varphi_{m_{1}},\chi_{m_{2}}\right\rangle \left\langle \varphi_{m_{1}},\chi_{m_{2}}\right\vert a_{\alpha}^{\dagger}\left\vert \varphi_{k_{1}},\chi_{k_{2}}\right\rangle \left\langle \varphi_{k_{1}},\chi_{k_{2}}\right\vert \rho(t)\left\vert \varphi_{n_{1}},\chi_{n_{2}}\right\rangle \nonumber \\
 & - & \left\langle \varphi_{n_{1}},\chi_{n_{2}}\right\vert \rho(t)\left\vert \varphi_{m_{1}},\chi_{m_{2}}\right\rangle \left\langle \varphi_{m_{1}},\chi_{m_{2}}\right\vert a_{\alpha}\left\vert \varphi_{k_{1}},\chi_{k_{2}}\right\rangle \left\langle \varphi_{k_{1}},\chi_{k_{2}}\right\vert a_{\alpha}^{\dagger}\left\vert \varphi_{n_{1}},\chi_{n_{2}}\right\rangle \Big\}.\label{3}
\end{eqnarray}
\end{small}
As the classical counterpart to the continuity Eq. (\ref{continuity})
we write the following master equation for two particles

\begin{equation}
\frac{\partial P_{n_{1}n_{2}}}{\partial t}=\sum_{m_{1},m_{2}}\left(T_{n_{1}n_{2}m_{1}m_{2}}P_{m_{1}m_{2}}-T_{m_{1}m_{2}n_{1}n_{2}}P_{n_{1}n_{2}}\right),
\end{equation}
where $T_{n_{1}n_{2}m_{1}m_{2}}dt$ is the transition probability
governing jumps from states $\left\vert \varphi_{n_{1}}\right\rangle $
and $|\chi_{m_{1}}\rangle$ to $\left\vert \varphi_{n_{2}}\right\rangle $
and $|\chi_{m_{2}}\rangle$, respectively. The quantum and stochastic
formalism meet a common ground through the mixed quantum-classical
equation
\begin{equation}
\frac{J_{n_{1}n_{2}m_{1}m_{2}}}{\hbar}=T_{n_{1}n_{2}m_{1}m_{2}}P_{m_{1}m_{2}}-T_{m_{1}m_{2}n_{1}n_{2}}P_{n_{1}n_{2}}\label{5}
\end{equation}
which admits the particular simplified solution 
\begin{equation}
T_{n_{1}n_{2}m_{1}m_{2}}=\left\{ \begin{array}{ccc}
\frac{J_{n_{1}n_{2}m_{1}m_{2}}}{\hbar P_{m_{1}m_{2}}} & , & J_{n_{1}n_{2}m_{1}m_{2}}\geq0\\
0 & , & J_{n_{1}n_{2}m_{1}m_{2}}\leq0
\end{array}\right.\text{.}\label{6}
\end{equation}

Next, we assume that the entangled systems are non-interacting harmonic
oscillators of frequencies $\omega_{\alpha}$ and masses $M_{\alpha}$,
each modeled by the Hamiltonian
\begin{equation}
H^{\left(\alpha\right)}=\frac{p_{\alpha}^{2}}{2M_{\alpha}}+\frac{M_{\alpha}\omega_{\alpha}^{2}}{2}x_{\alpha}^{2},\label{7}
\end{equation}
$p_{\alpha}$ being the canonically conjugate momentum to the coordinate
variable $x_{\alpha}$, and the total Hamiltonian is $H=H^{(1)}+H^{(2)}$.
Additionally, we consider both entangled systems to be described by
the arbitrary general mixed state 
\begin{equation}
\rho(t)=\sum_{u}P_{u}\left\vert \psi^{u}(t)\right\rangle \left\langle \psi^{u}(t)\right\vert ,\label{8}
\end{equation}
 with $\left\langle \varphi_{n_{1}},\chi_{n_{2}}\right.\left\vert \psi^{u}(t)\right\rangle \equiv\psi_{n_{1},n_{2}}^{u}$.

In Vink's extension of Bell's beables \cite{Vink}, where all the
degrees of freedom must be discrete and finite, the position is restricted
to sites of a lattice which, in the one-dimensional case, becomes
$x_{n_{\alpha}}=n_{\alpha}\varepsilon$, $n_{\alpha}$ being integers
and $\varepsilon$ is the lattice distance. To extend Vink's approach
to two particles (the continuous limit is recovered taking $\varepsilon\rightarrow0$)
we must \emph{i}) write the smooth wave functions in the coordinate
representations as $\left(\left\vert \varphi_{n_{\alpha}}\right\rangle =\left\vert x_{n_{\alpha}}\right\rangle \right)$
$\psi_{n_{1},n_{2}}^{u}=R_{n_{1},n_{2}}^{u}\exp\left[\frac{i}{\hbar}S_{n_{1},n_{2}}^{u}\right]$,
where $\psi_{n_{1},n_{2}}^{u}\equiv\psi^{u}\left(x_{n_{1}},x_{n_{2}},t\right)$,
$R_{n_{1},n_{2}}^{u}\equiv R^{u}\left(x_{n_{1}},x_{n_{2}},t\right)$,
and $S_{n_{1},n_{2}}^{u}\equiv S^{u}\left(x_{n_{1}},x_{n_{2}},t\right)$;\emph{
ii)} expand $\psi^{u}$ to first order in $\varepsilon$, \emph{i.e.,}
\begin{align}
\psi_{n_{1}\pm1,n_{2}}^{u} & =\psi_{n_{1},n_{2}}^{u}\pm\varepsilon\triangle_{1}\psi_{n_{1},n_{2}}^{u}\label{9}\\
\psi_{n_{1},n_{2}\pm1}^{u} & =\psi_{n_{1},n_{2}}^{u}\pm\varepsilon\triangle_{2}\psi_{n_{1},n_{2}}^{u}
\end{align}
with
\begin{equation}
\Delta_{\alpha}\psi_{n_{1},n_{2}}^{u}=\left[\Delta_{\alpha}R_{n_{1},n_{2}}^{u}+\frac{i}{\hbar}R_{n_{1},n_{2}}^{u}\Delta_{\alpha}S_{n_{1},n_{2}}^{u}\right]\exp\left(\frac{i}{\hbar}S_{n_{1},n_{2}}^{u}\right)\text{;}\label{10}
\end{equation}
\emph{iii}) substitute $a_{n_{\alpha}m_{\alpha}}=[M_{\alpha}\omega_{\alpha}x_{m_{\alpha}}\delta_{n_{\alpha},m_{\alpha}}+\hbar(\delta_{n_{\alpha}+1,m_{\alpha}}-\delta_{n_{\alpha},m_{\alpha}})/\varepsilon]/(2\hbar M_{\alpha}\omega_{\alpha})^{1/2}$in
the transition matrix given by $J_{n_{1}n_{2}m_{1}m_{2}}/\hbar$,
neglect terms of order O($\epsilon^{2}$) and higher (taking $\varepsilon_{\alpha}=\varepsilon$
for simplicity); \emph{iv}) take the limit $\epsilon\rightarrow0$
with $x_{m_{2}\pm1}=\varepsilon m_{2}\pm\varepsilon$. After a straightforward
but length calculations we obtain ($\alpha,\beta=1,2$):
\begin{small}
\begin{equation}
\frac{J_{n_{1}n_{2}m_{1}m_{2}}}{\hbar}=\sum_{u}P_{u}\left(t\right)R_{u}^{2}\sum_{\left(\alpha\neq\beta\right)}\left[\frac{1}{M_{\alpha}}\Delta_{\alpha}S_{u}-\frac{\hbar\gamma_{\alpha}\left(2\bar{n}_{\alpha}+1\right)}{2M_{\alpha}\omega_{\alpha}}\frac{1}{R_{u}}\Delta_{\alpha}R_{u}\right]\left(\delta_{m_{\alpha}+1,n_{\alpha}}-\delta_{m_{\alpha}-1,n_{\alpha}}\right)\delta_{m_{\beta},n_{\beta}}.
\end{equation}
\end{small}
Next, by defining
\begin{equation}
x_{\alpha}\left(t+dt\right)\simeq x_{\alpha}\left(t\right)+\varepsilon[\langle k-m\rangle\delta_{\alpha,1}+\langle l-n\rangle\delta_{\alpha,2}],\label{13}
\end{equation}
with $\langle k-m\rangle=\sum_{mn}T_{mknl}\left(k-m\right)dt$ and
$\left\langle l-n\right\rangle =\sum_{mn}T_{mknl}\left(l-n\right)dt$,
such that for the forward movement $n_{1}-m_{1}=1$ \emph{i.e}., $k>m$
($l>n$) for particle $1$ (2), Eq. (\ref{13}) results in 
\begin{align*}
\frac{dx_{\alpha}\left(t\right)}{dt} & =\frac{1}{\sum_{u}P_{u}\left(R^{u}(x_{1},x_{2},t)\right)^{2}}\\
\\
\times & \sum_{u}\left\{ P_{u}\left(R^{u}(x_{1},x_{2},t)\right)^{2}\left[\frac{1}{M_{\alpha}}\frac{\partial S^{u}(x_{1},x_{2},t)}{\partial x_{\alpha}}+\frac{\hbar\gamma_{\alpha}\left(2\bar{n}_{\alpha}+1\right)}{2M_{\alpha}\omega_{\alpha}}\left(\frac{1}{R^{u}(x_{1},x_{2},t)}\frac{\partial R^{u}(x_{1},x_{2},t)}{\partial x_{\alpha}}\right)\right]\right\} \text{.}\\
\end{align*}

The above equation can be rewritten more compactly in the following
way, using $\rho\left(x_{1},x_{2},x_{1}^{\prime},x_{2}^{\prime},t\right)=\left\langle x_{1},x_{2}\right\vert \rho\left(t\right)\left\vert x_{1}^{\prime},x_{2}^{\prime}\right\rangle $,
as 
\begin{equation}
\frac{dx_{\alpha}\left(t\right)}{dt}=\frac{\hbar}{M_{\alpha}}\left[\frac{{\rm {\rm Im}}\left[\partial_{x_{\alpha}}\rho\left(x_{1},x_{2},x_{1}^{\prime},x_{2}^{\prime},t\right)\right]}{\rho\left(x_{1},x_{2},x_{1}^{\prime},x_{2}^{\prime},t\right)}\right]_{\substack{x_{1}=x_{1}^{\prime}\\
x_{2}=x_{2}^{\prime}
}
}+\frac{\hbar\gamma_{\alpha}\left(2\bar{n}_{\alpha}+1\right)}{2M_{\alpha}\omega_{\alpha}}\left[\frac{{\rm Re}\left[\partial_{x_{\alpha}}\rho\left(x_{1},x_{2},x_{1}^{\prime},x_{2}^{\prime},t\right)\right]}{\rho\left(x_{1},x_{2},x_{1}^{\prime},x_{2}^{\prime},t\right)}\right]_{\substack{x_{1}=x_{1}^{\prime}\\
x_{2}=x_{2}^{\prime}
}
}\text{.}\label{13-1}
\end{equation}

We note that our main result, Eq. (\ref{13-1}), generalizes that
one obtained by a different approach in Ref.\cite{Maroney} for density
matrix of individual systems without dissipation ($\gamma_{\alpha}=0$).
In the following, we shall use Eq. (\ref{13-1}) to calculate trajectories
of quantum particles when the entanglement either is present or absent
in the joint state.

\section{generalized Werner states and motion equations}

With the Bohmian equation of motion for two particles in hands, Eq.
(\ref{13-1}), we next assume that the entangled state is prepared
in the generalized Werner state
\begin{equation}
\rho=\epsilon\left\vert \psi^{\pm}\right\rangle _{12}\left\langle \psi^{\pm}\right\vert +\frac{1-\epsilon}{4}\mathbb{I}\text{,}\label{14}
\end{equation}
where $\mathbb{I}\equiv\mathbb{I}_{1}\otimes\mathbb{I}_{2}$ stands
for the identity operator and 
\begin{equation}
\left\vert \psi^{\pm}\right\rangle _{12}=a\left\vert 00\right\rangle \pm b\left\vert 11\right\rangle ,\label{15}
\end{equation}
with $a$ and $b$ being complex constants ($|a|^{2}+|b|^{2}=1$).

To compute Bohmian trajectories we must find the solution $\rho(t)$
in the presence of losses due to a thermal reservoir. We then will
specialize to the case of losses at zero temperature, using the method
of \emph{phenomenological operator approach, }as developed in Ref.\cite{POA},
where we define ($\gamma_{\alpha}=\gamma$)

\begin{eqnarray*}
\left\vert 0\right\rangle \left\vert {\normalcolor 0}\right\rangle _{R} & \rightarrow & \left\vert 0\right\rangle \left\vert 0\right\rangle _{R}\\
\left\vert 1\right\rangle \left\vert 0\right\rangle _{R} & \rightarrow & e^{-\frac{\gamma}{2}t}\left\vert 1\right\rangle \left\vert 0\right\rangle _{R}+\sqrt{1-e^{-\gamma t}}\left\vert 0\right\rangle \left\vert 1\right\rangle _{R}.
\end{eqnarray*}
 If we now take into account that the Werner-like state is uncoupled
from the reservoir at $t=0$, then using the rules given above we
can write, after tracing out the reservoir variables,

\begin{eqnarray}
\rho\left(t\right) & = & \left\{ \left[\epsilon a^{2}+\frac{1-\epsilon}{4}\right]+\left[\epsilon b^{2}+\frac{1-\epsilon}{4}\right]\left(1-e^{-\gamma t}\right)^{2}+\frac{1-\epsilon}{2}\left(1-e^{-\gamma t}\right)\right\} \left\vert 00\right\rangle \left\langle 00\right\vert \nonumber \\
 &  & +\left[\epsilon b^{2}+\frac{1-\epsilon}{4}\right]e^{-2\gamma t}\left\vert 11\right\rangle \left\langle 11\right\vert \nonumber \\
 &  & +e^{-\gamma t}\left\{ \left[\epsilon b^{2}+\frac{1-\epsilon}{4}\right]\left(1-e^{-\gamma t}\right)+\frac{1-\epsilon}{4}\right\} \left(\left\vert 10\right\rangle \left\langle 10\right\vert +\left\vert 01\right\rangle \left\langle 01\right\vert \right)\nonumber \\
 &  & \pm\epsilon e^{-\gamma t}\left[ab^{\ast}\ e^{2i\omega t}\left\vert 00\right\rangle \left\langle 11\right\vert +ba^{\ast}e^{-2i\omega t}\left\vert 11\right\rangle \left\langle 00\right\vert \right],\label{eq:16}
\end{eqnarray}
where we have chosen $\omega_{\alpha}=\omega$. We now assume that
both entangled particles, originally represented in abstract Fock
spaces, are harmonic oscillators within the subspace $\left\{ 0,1\right\} $
of the ground and first excited states. This assumption enables us
to analyze, through both particles' trajectories, how entanglement
dynamics affects Bohmian trajectories. Considering the scaled dimensionless
variables $\tilde{x}_{\alpha}=\sqrt{\omega/\hbar}x_{\alpha}$, we
write the state (\ref{eq:16}) in the coordinate representation to
obtain $\rho\left(x_{1},x_{2},x_{1}^{\prime},x_{2}^{\prime},t\right)$
as
\begin{align}
\rho\left(x_{1},x_{2},x_{1}^{\prime},x_{2}^{\prime},t\right) & =\left(\frac{\omega}{\pi\hbar}\right)\Big\{\left[\epsilon a^{2}+\frac{1-\epsilon}{4}\right]+\left[\epsilon b^{2}+\frac{1-\epsilon}{4}\right]\left(1-e^{-\gamma t}\right)^{2}+\frac{1-\epsilon}{2}\left(1-e^{-\gamma t}\right)\nonumber \\
 & +4\left[\epsilon b^{2}+\frac{1-\epsilon}{4}\right]e^{-2\gamma t}\tilde{x_{1}}\tilde{x_{1}}^{\prime}\tilde{x_{2}}\tilde{x_{2}}^{\prime}\nonumber \\
 & +2e^{-\gamma t}\left\{ \left[\epsilon b^{2}+\frac{1-\epsilon}{4}\right]\left(1-e^{-\gamma t}\right)+\frac{1-\epsilon}{4}\right\} \left[\tilde{x_{1}}\tilde{x_{1}}^{\prime}\tilde{+x_{2}}\tilde{x_{2}}^{\prime}\right]\nonumber \\
 & \pm2\epsilon e^{-\gamma t}\left[ab^{\ast}\ e^{2i\omega t}\tilde{x}_{1}^{\prime}\tilde{x_{2}}^{\prime}+ba^{\ast}e^{-2i\omega t}\tilde{x}_{1}\tilde{x_{2}}\right]\Big\}\times e^{-\frac{1}{2}\left[\tilde{x_{1}}^{2}+\left(\tilde{x_{1}}^{\prime}\right)^{2}+\tilde{x_{2}}+\left(\tilde{x_{2}}^{\prime}\right)^{2}\right]}{}^{.}\label{23}
\end{align}
Using Eq. (\ref{13-1}) we obtain, after a straightforward calculation

\begin{eqnarray}
\frac{d\tilde{x}_{1}}{dt} & = & \frac{\gamma}{\mathcal{G}\left(\tilde{x_{1}},\tilde{x_{2}};t\right)}\left[\mp\frac{2\omega}{\gamma}A(t)\sin(2\omega t)\tilde{x_{2}}+B(t)\tilde{x_{1}}\tilde{x_{2}}^{2}+C(t)\tilde{x_{1}}\pm A(t)\cos(2\omega t)\tilde{x_{2}}\right]-\frac{\gamma\tilde{x_{1}}}{2},\label{xponto}\\
\frac{d\tilde{x}_{2}}{dt} & = & \frac{\gamma}{\mathcal{G}\left(\tilde{x_{1}},\tilde{x_{2}};t\right)}\left[\mp\frac{2\omega}{\gamma}A(t)\sin(2\omega t)\tilde{x_{1}}+B(t)\tilde{x_{1}}^{2}\tilde{x_{2}}+C(t)\tilde{x_{2}}\pm A(t)\cos(2\omega t)\tilde{x_{1}}\right]-\frac{\gamma\tilde{x_{2}}}{2},\label{yponto}
\end{eqnarray}

where

\begin{eqnarray}
\mathcal{G}\left(\tilde{x_{1}},\tilde{x_{2}};t\right) & = & \left[\epsilon a^{2}+\frac{1-\epsilon}{4}\right]+\left[\epsilon b^{2}+\frac{1-\epsilon}{4}\right]\left(1-e^{-\gamma t}\right)^{2}\nonumber \\
 & + & \frac{1-\epsilon}{2}\left(1-e^{-\gamma t}\right)+4\left[\epsilon b^{2}+\frac{1-\epsilon}{4}\right]e^{-2\gamma t}\tilde{x_{1}}^{2}\tilde{x_{2}}^{2}\nonumber \\
 & + & 2e^{-\gamma t}\left\{ \left[\epsilon b^{2}+\frac{1-\epsilon}{4}\right]\left(1-e^{-\gamma t}\right)+\frac{1-\epsilon}{4}\right\} \left[\tilde{x_{1}}^{2}+\tilde{x_{2}}^{2}\right]\nonumber \\
 & \pm & 4\epsilon ab\tilde{x_{1}}\tilde{x_{2}}e^{-\gamma t}\cos(2\omega t),\label{eq:22}
\end{eqnarray}

\begin{eqnarray}
A(t) & = & \epsilon abe^{-\gamma t},\label{eq:23}
\end{eqnarray}

\begin{eqnarray}
B(t) & = & 2\left[\epsilon b^{2}+\frac{1-\epsilon}{4}\right]e^{-2\gamma t},\label{eq:24}
\end{eqnarray}

\begin{equation}
C(t)=2e^{-\gamma t}\left\{ \left[\epsilon b^{2}+\frac{1-\epsilon}{4}\right]\left(1-e^{-\gamma t}\right)+\frac{1-\epsilon}{4}\right\} .
\end{equation}

In the next section we will explore these solutions plotting the corresponding
quantum trajectories for the generalized Werner state of Eq. (\ref{14})
in regions occurring entanglement or separability.

\section{Quantum trajectories for entangled states}

In this section we present our results regarding Bohmian trajectories
for both separable and entangled states using the generalized Werner
state given in Eq. (\ref{14}). To quantify the entanglement present
in this state we can either use the concurrence \cite{Wooters98}
or the negativity \cite{Peres96,Hayden01}, thus we will use the concurrence
as defined for two-qubit system:
\begin{equation}
C(\rho)={\rm max}\left\{ 0,\sqrt{\lambda_{1}}-\sqrt{\lambda_{2}}-\sqrt{\lambda_{3}}-\sqrt{\lambda_{4}}\right\} ,\label{2-2-1}
\end{equation}
where $\lambda_{k}$ are the eigenvalues of the matrix $\tilde{\rho}_{12}\left(t\right)=\sigma_{y}^{1}\sigma_{y}^{2}\rho_{12}^{*}(t)\sigma_{y}^{1}\sigma_{y}^{2}$
arranged in decreasing order. 

As is well known, when disregarding losses and\emph{ $a=b$, }this
state is separable for\emph{ $\epsilon=1/3$}. In Figs. 1(a) and 1(b)
we show the Bohmian trajectories for the generalized Werner state
of Eq. (\ref{14}) for \emph{$a=b$ }and $\epsilon=0.1$, $1/3$,
corresponding to a separable state. In Figs. 2(a) and 2(b) we show
the quantum trajectories when \emph{$a=b$ }and $\epsilon=0.4$, $1$,
corresponding to an entangled state. From this sample of figures it
can be seen that for separable state ($\epsilon<1/3$) the amplitude
of oscillations of each trajectory is smooth and relatively small
as compared with the corresponding trajectory (same initial condition)
for entangled states ($\epsilon>1/3$), as advanced in Fig. 5(a) where
we have plotted $x(t)$ \textit{versus }$\epsilon$ for the same initial
condition. From Figs. 2(a) and 2(b) we observe that, as the mixing
parameter ($\epsilon$) increases, there is a corresponding squeezing
of the trajectories in regions where both particles approach each
other. This squeeze of trajectories, which clearly increases from
Fig. 2(a) to Fig. 2(b), together with the increasing oscillation amplitude,
works as a signature of entanglement for the generalized Werner state
studied here.

\begin{figure}[tb]
\centering
\includegraphics[width=7cm]{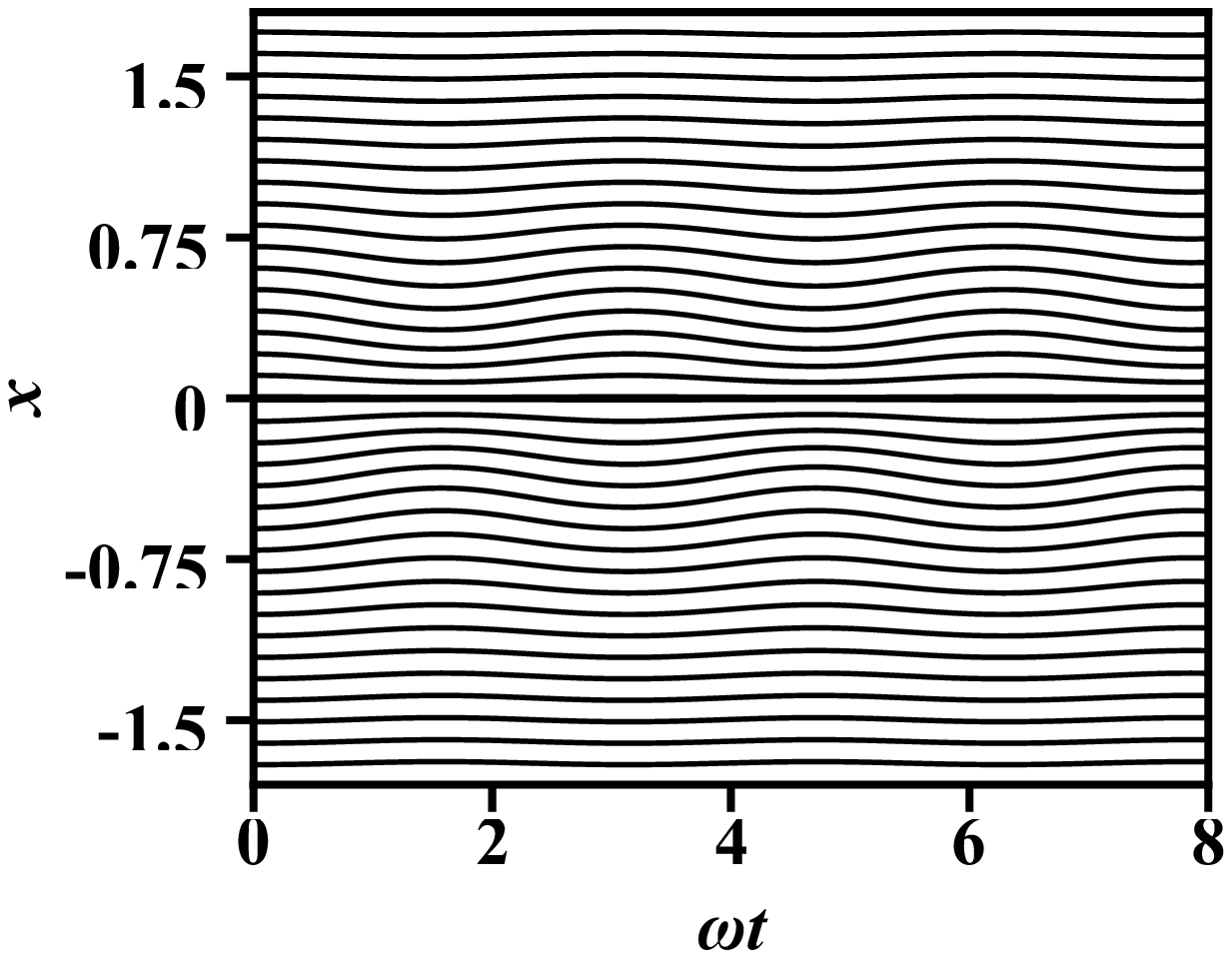}\hfil
\includegraphics[width=7cm]{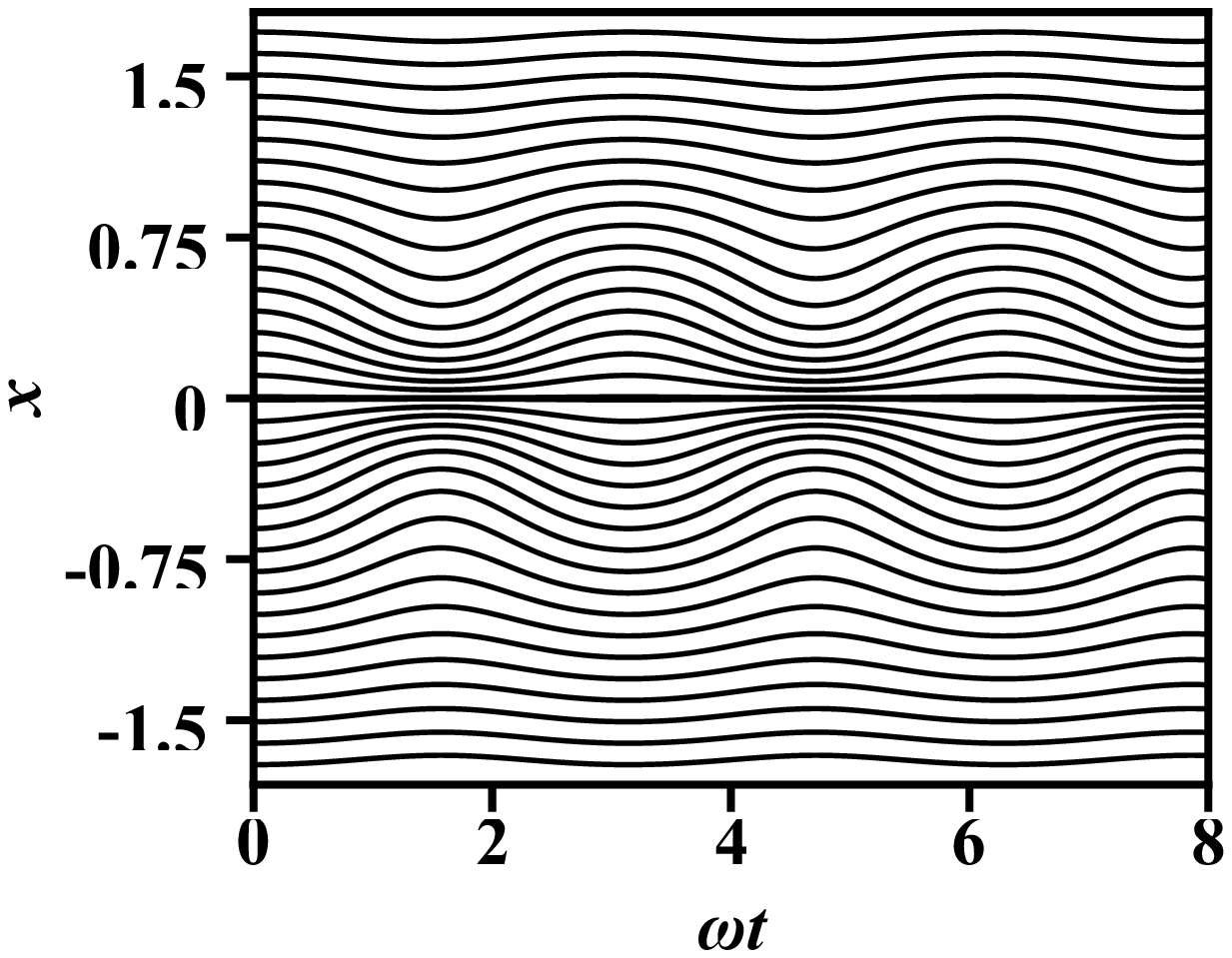}
\caption{Bohmian trajectories for the generalized Werner state of Eq. (\ref{14})
with $a=b$. For separable states with (a) $\epsilon=0.1$ and (b)
$\epsilon=1/3$.}
\end{figure}

\begin{figure}[tb]
\centering
\includegraphics[width=7cm]{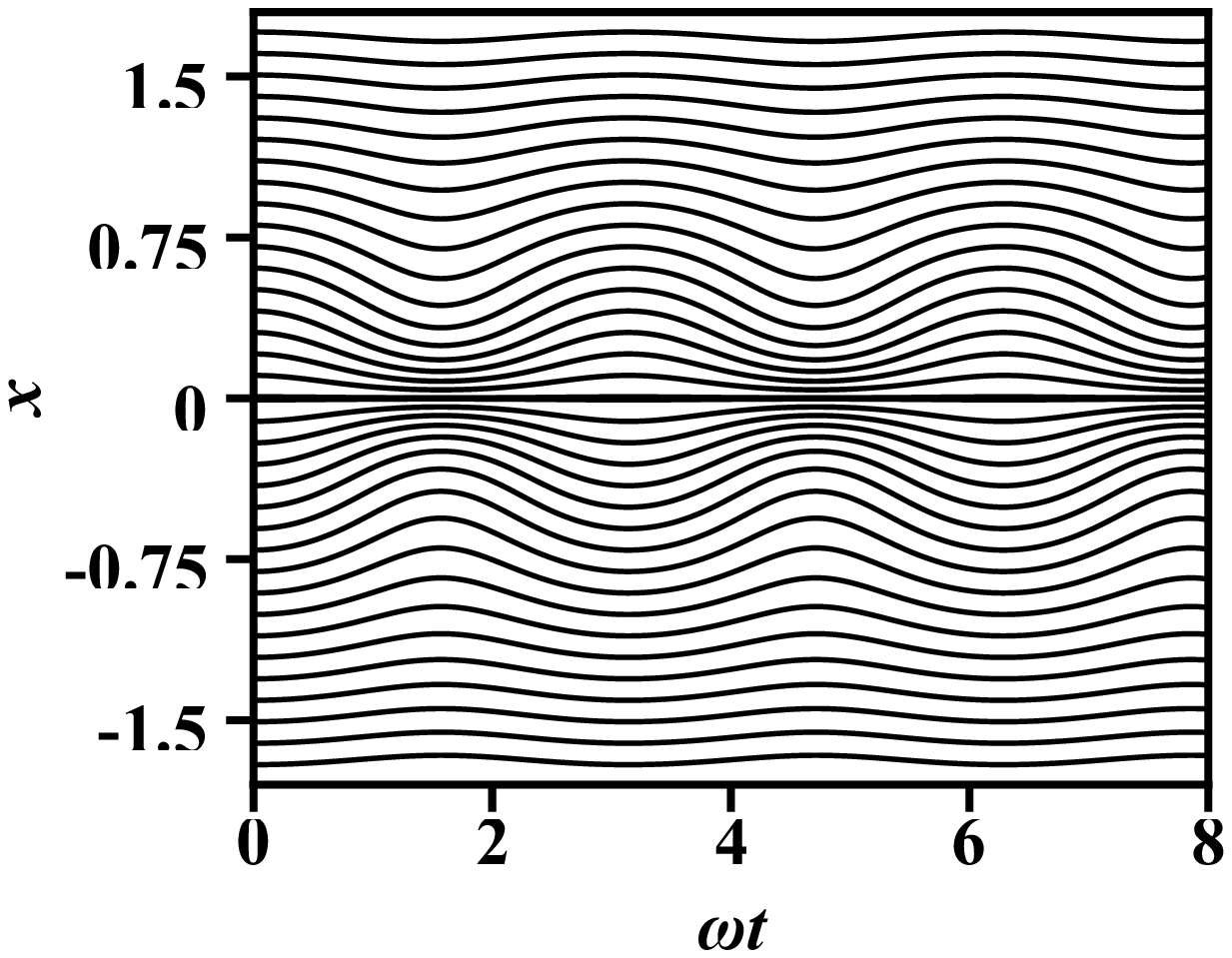}\hfil
\includegraphics[width=7cm]{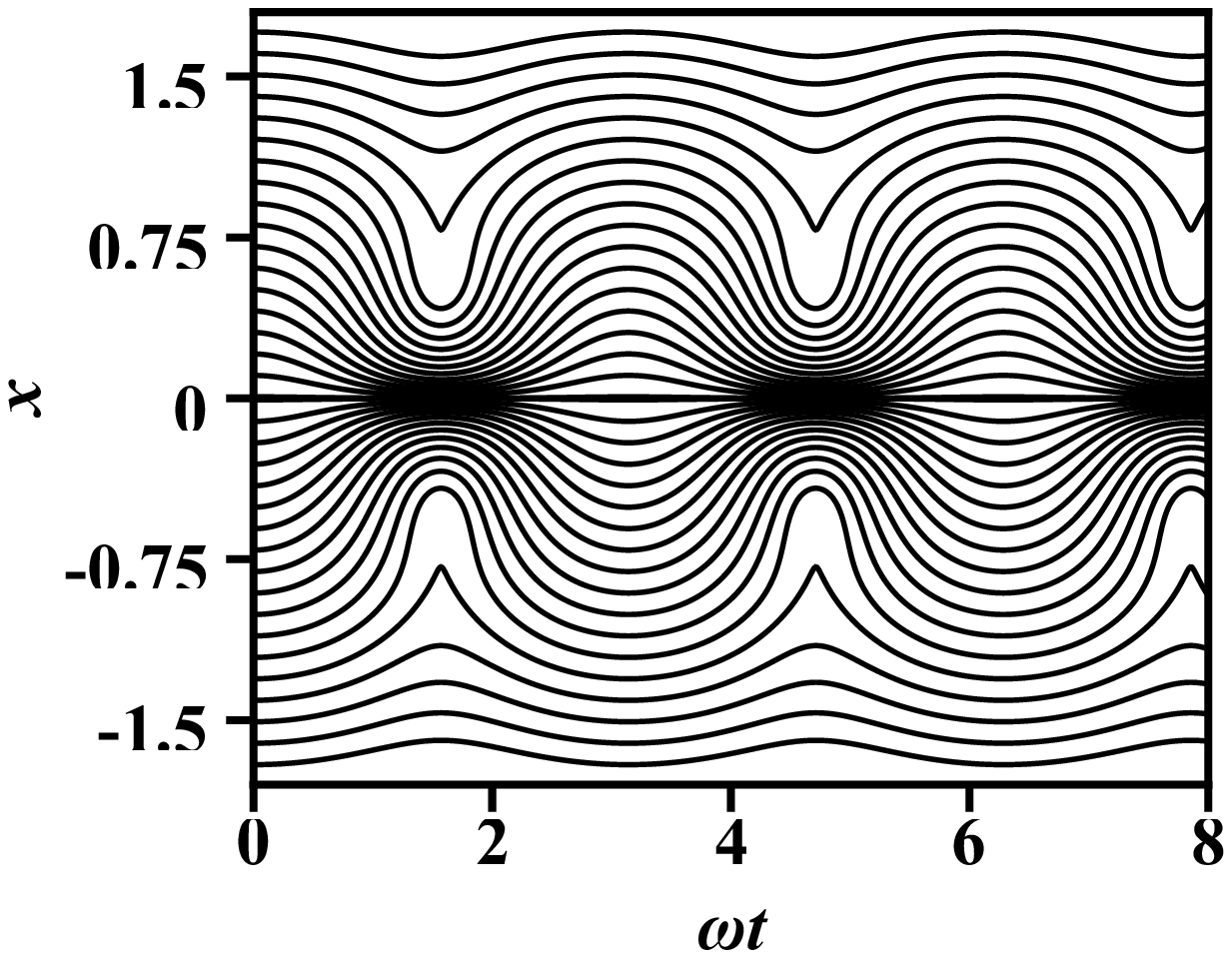}
\caption{Bohmian trajectories for the generalized Werner state of Eq.(\ref{14})
with $a=b$. For entangled states with (a) $\epsilon=0.4$ and (b)
$\epsilon=1$.}
\end{figure}

\begin{figure}[tb]
\centering
\includegraphics[width=7cm]{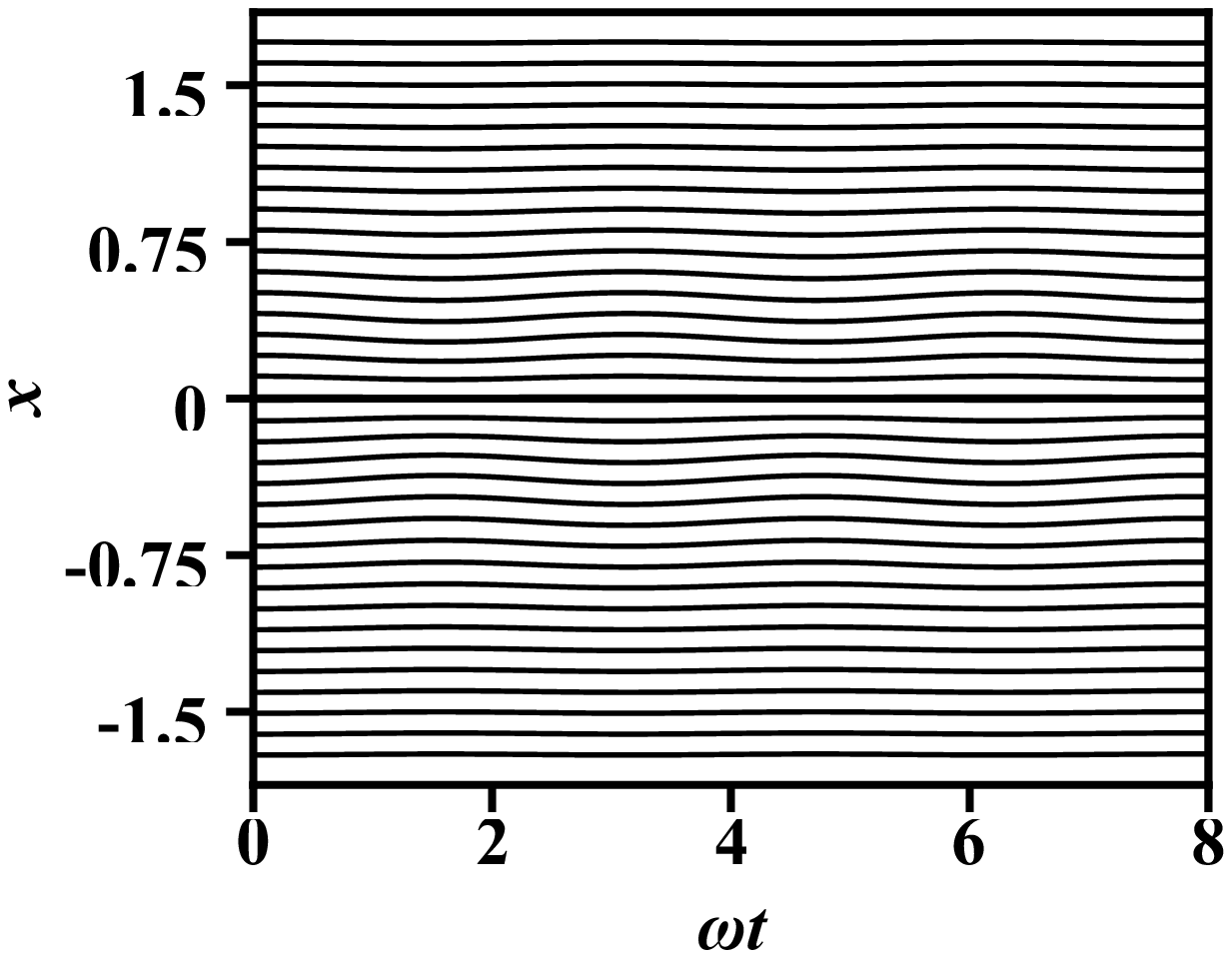}\hfil
\includegraphics[width=7cm]{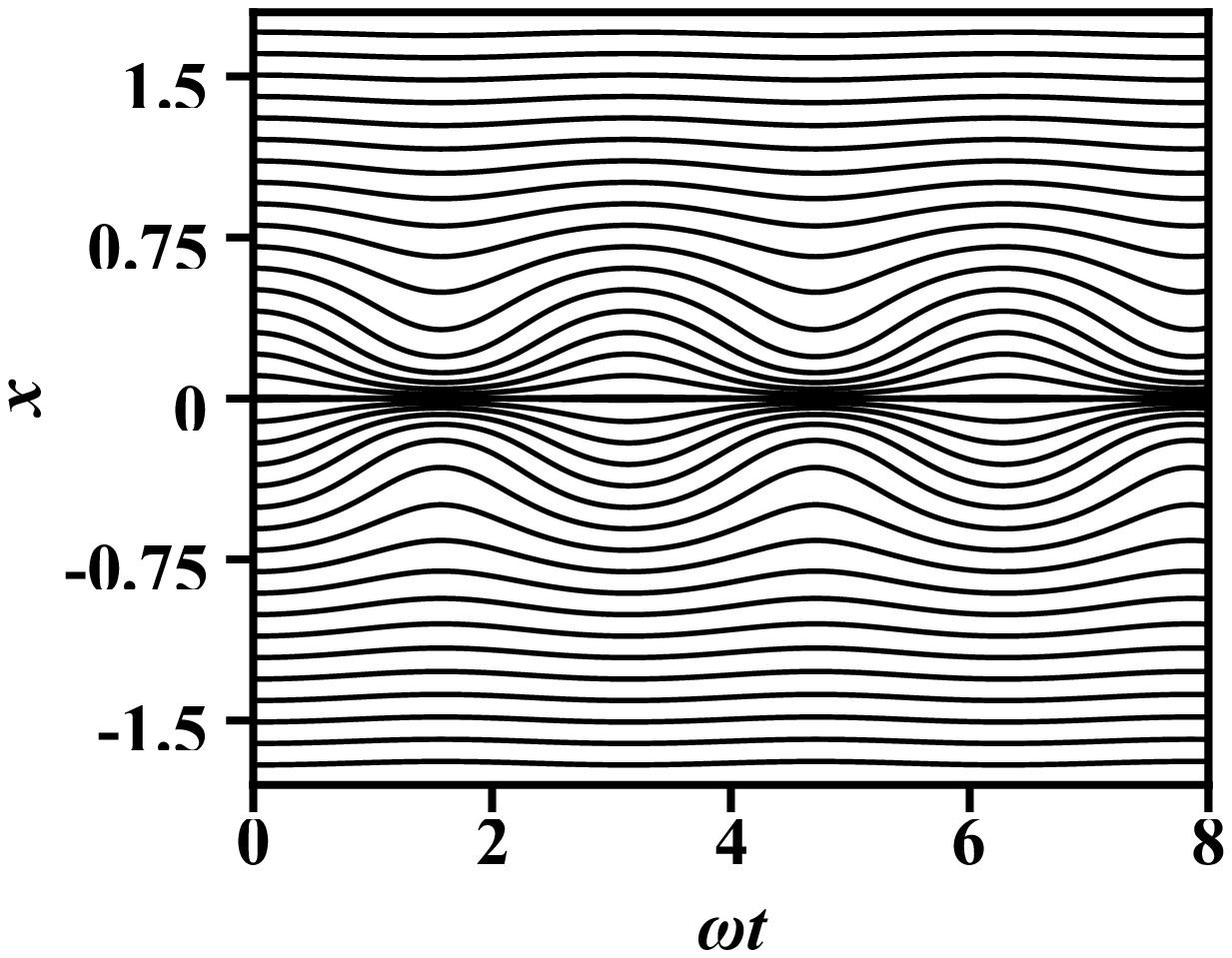}
\caption{Bohmian trajectories for the generalized Werner state of Eq. (\ref{14})
with $a=0.2$. Entanglement occurs for $\epsilon\geq0.56$. (a) $\epsilon=0.1$
and (b) $\epsilon=0.56$.}
\end{figure}

\begin{figure}[tb]
\centering
\includegraphics[width=7cm]{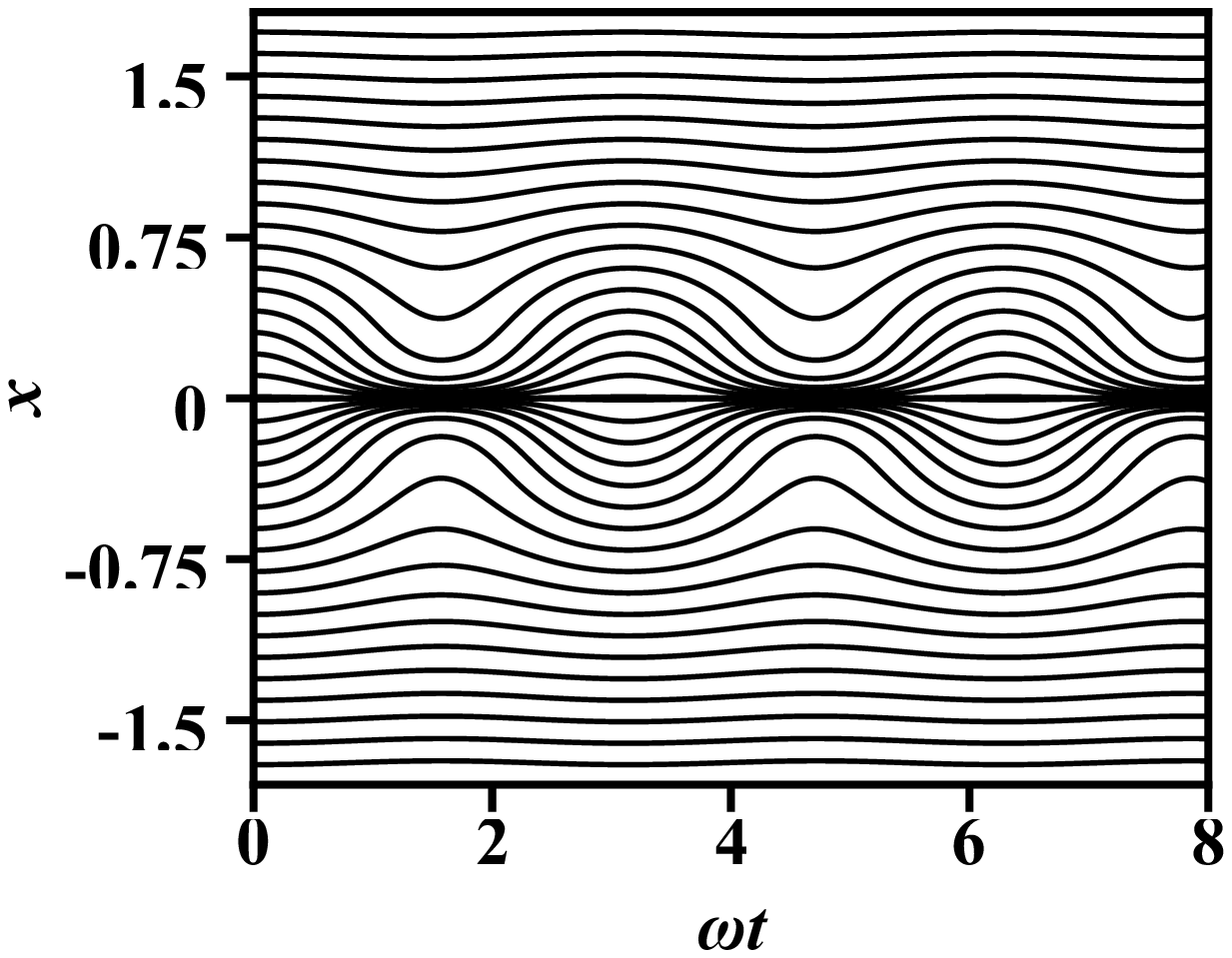}\hfil
\includegraphics[width=7cm]{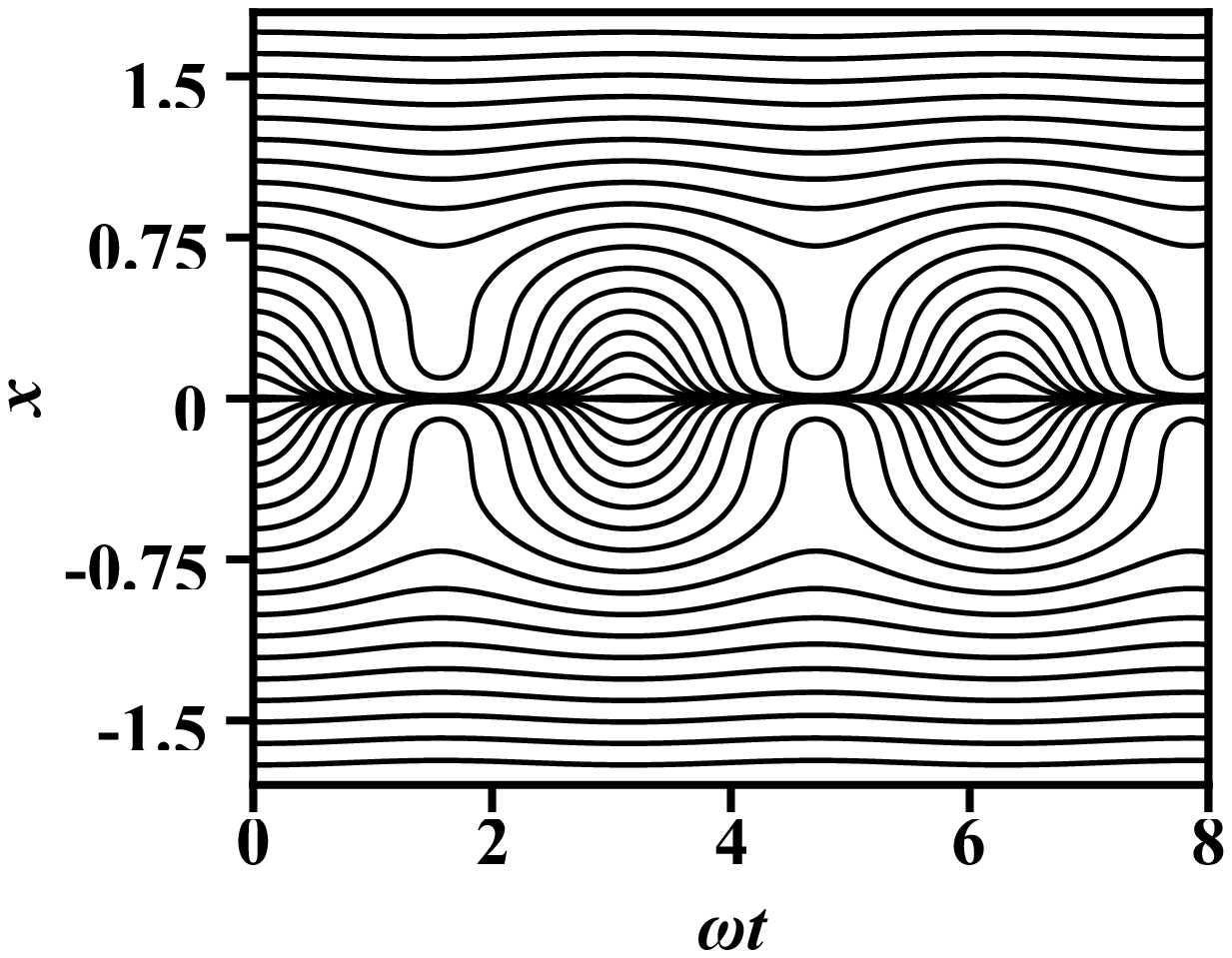}
\caption{Bohmian trajectories for the generalized Werner state of Eq. (\ref{14})
with $a=0.2$. Entanglement occurs for $\epsilon\geq0.56$. (a) $\epsilon=0.7$
(b) and $\epsilon=1.0$.}
\end{figure}

\begin{figure}[tb]
\centering
\includegraphics[width=7cm]{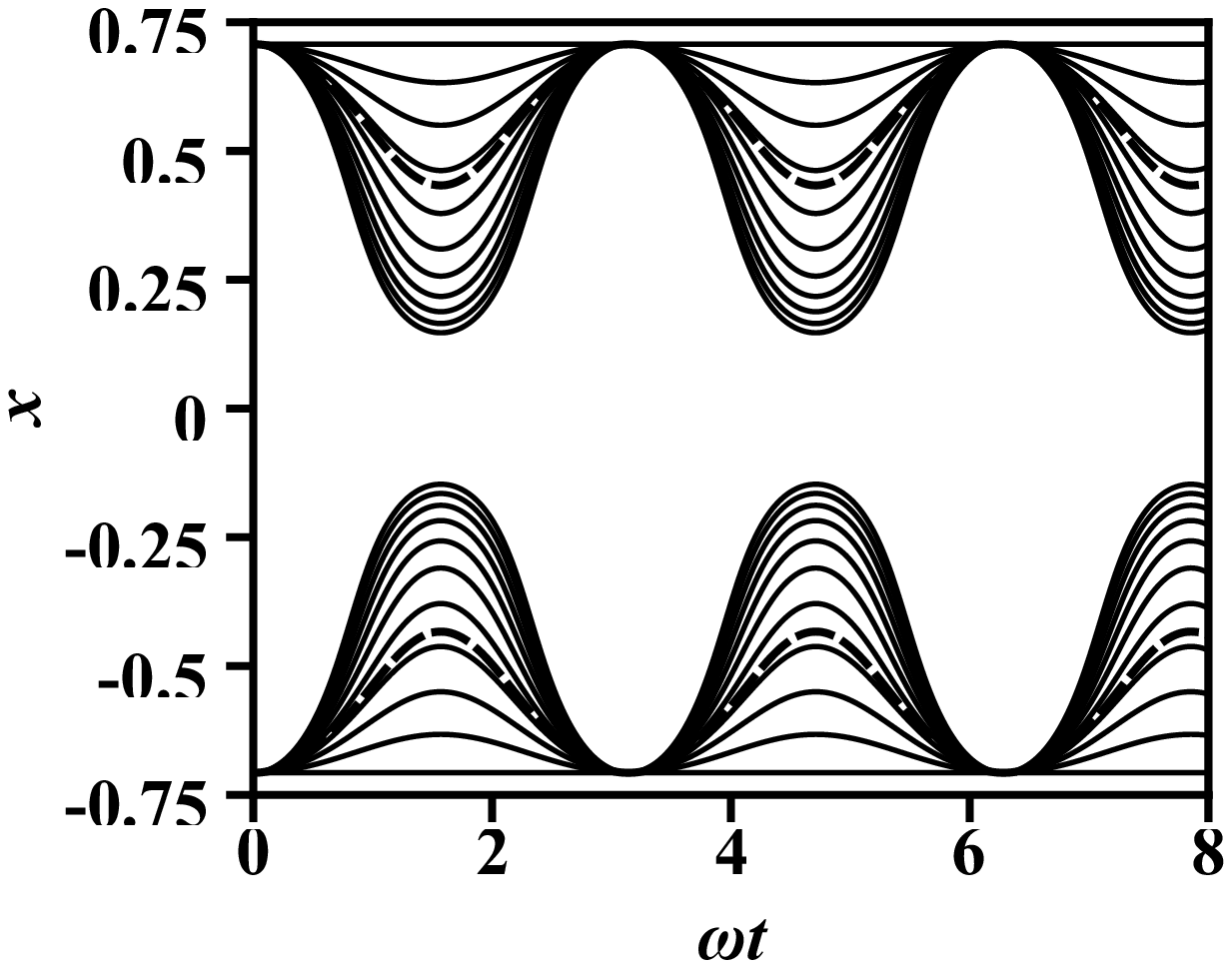}\hfil
\includegraphics[width=7cm]{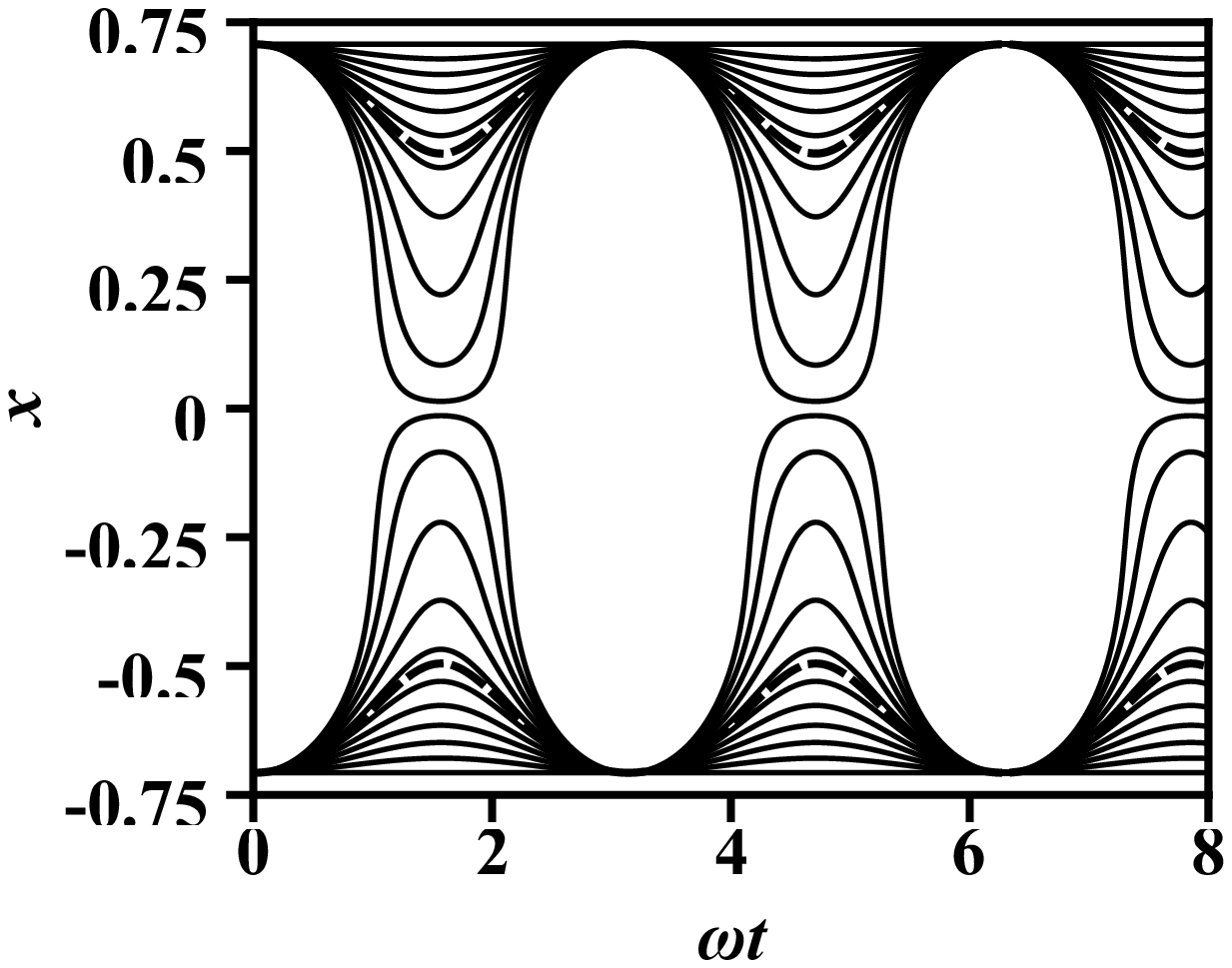}
\caption{Bohmian trajectories for (a) $a=b$ and (b) $a=0.2$, for the same
initial conditions and several values of $\epsilon$. The constant
trajectory is for $\epsilon=0$, while the oscilation amplitude is
maximum for $\epsilon=1.0$. The trajectories for (a) $\epsilon=1/3$
and (b) $\varepsilon=0.56$ are indicated by a dashed lines.}
\end{figure}

\begin{figure}[tb]
\centering
\includegraphics[width=7cm]{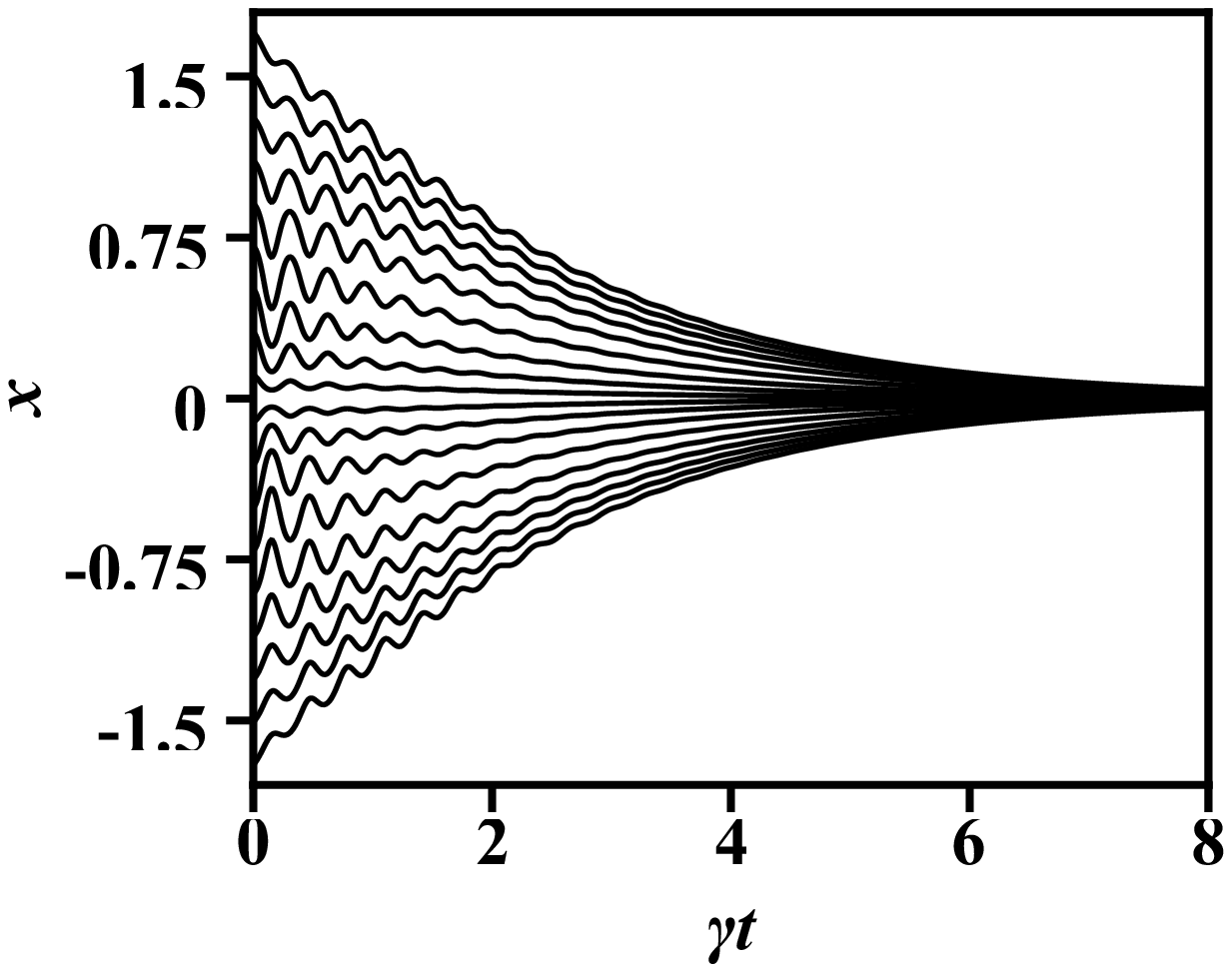}\hfil
\includegraphics[width=7cm]{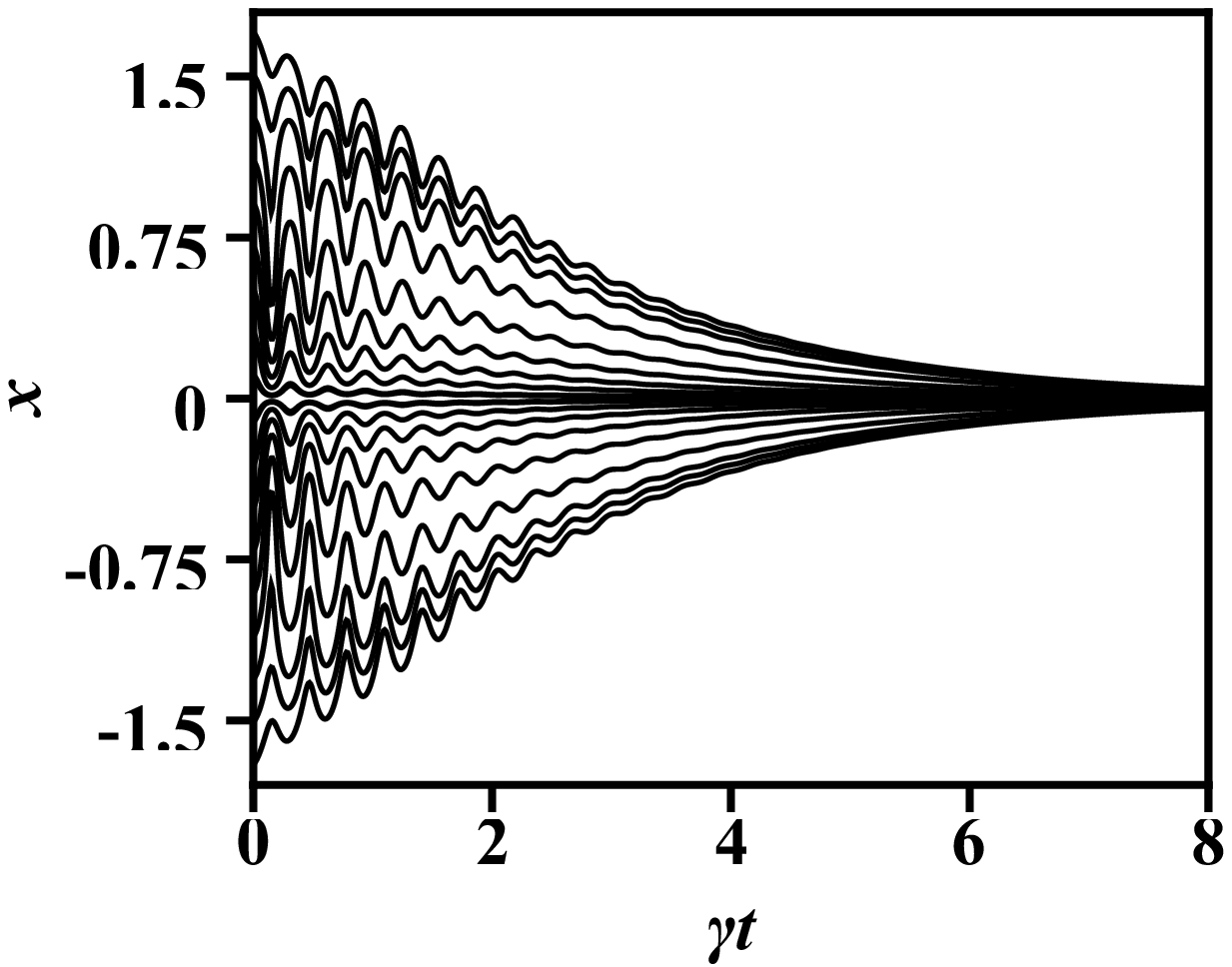}
\caption{Bohmian trajectories for the generalized Werner state of Eq. (\ref{14})
with $a=b$, $\omega/\gamma=10.0$ (a) for $\epsilon=0.4$, where
entanglement is suddenly lost at $\gamma t=0.15$, and (b) for $\epsilon=1.0$,
where entanglement is lost at the asymptotic time.}
\end{figure}

\begin{figure}[b]
\centering
\includegraphics[width=7cm]{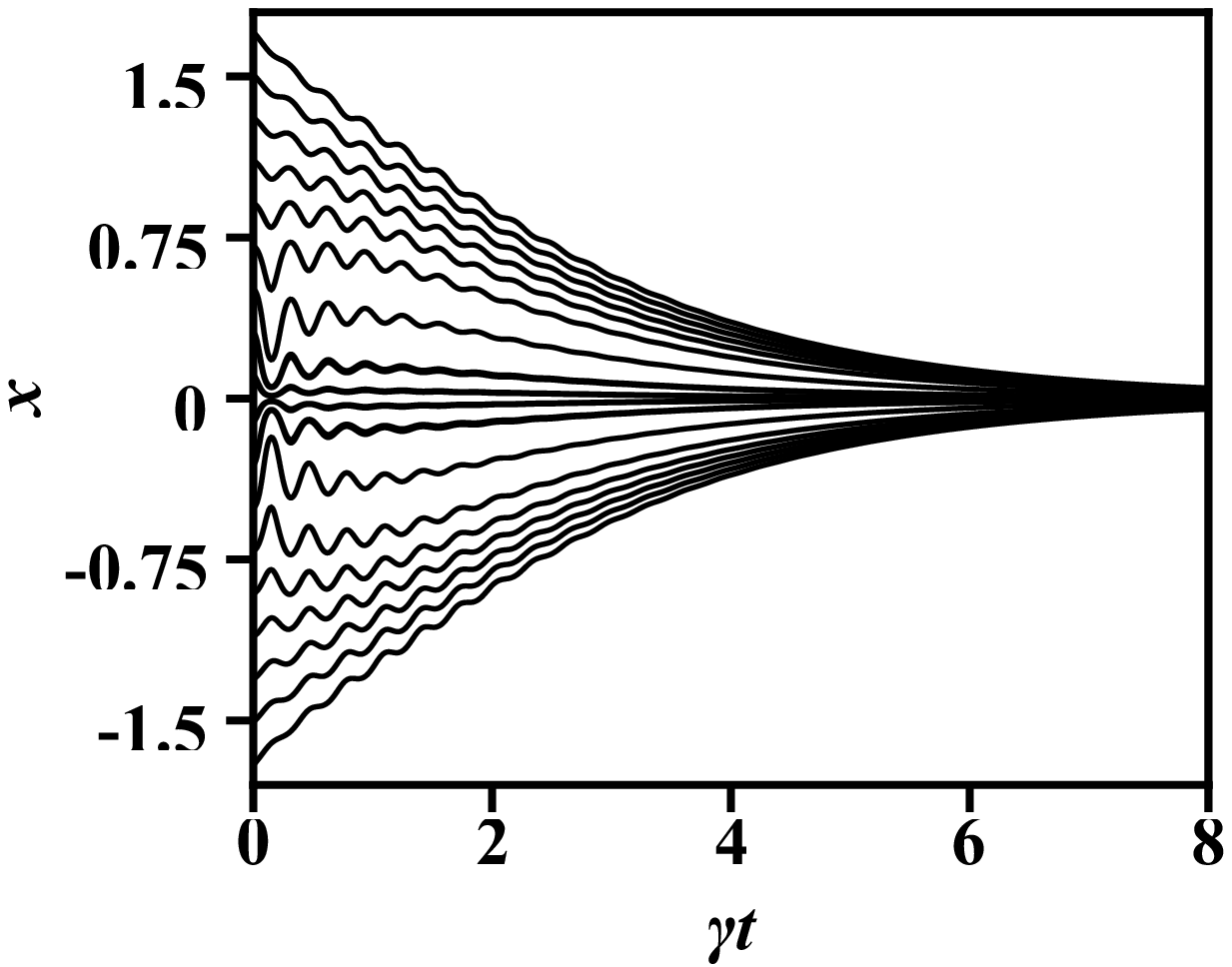}\hfil
\includegraphics[width=7cm]{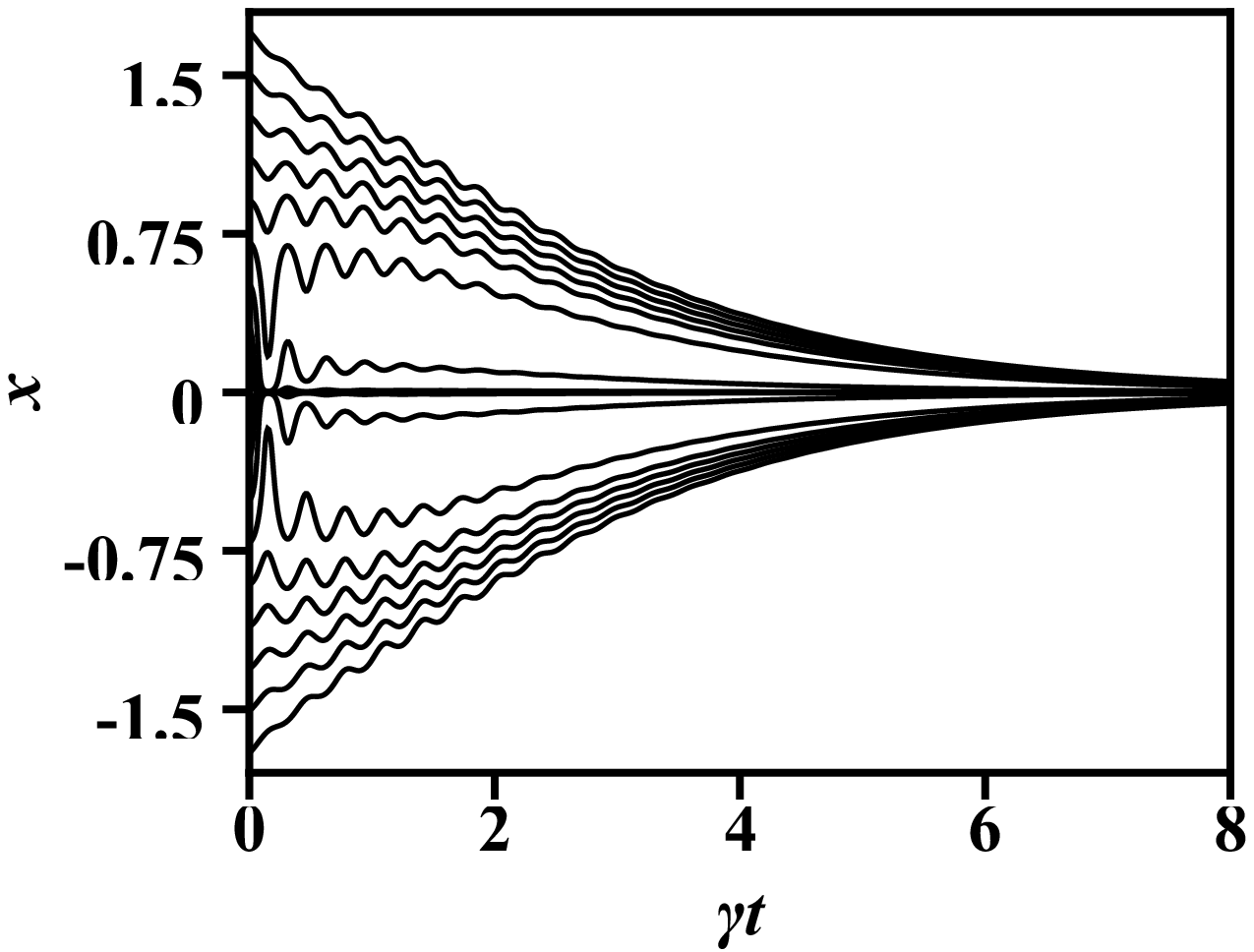}
\caption{Bohmian trajectories for the generalized Werner state of Eq. (\ref{14})
with $a=0.2$, and $\omega/\gamma=10$.0 for (a) for $\epsilon=0.7,$
where entanglement is suddenly lost at $\gamma t=0.026$ and (b) $\epsilon=1.0,$
where entanglement is lost at $\gamma t=0.23$. }
\end{figure}

Using Eq. (\ref{2-2-1}), we can readily check that disregarding losses
and considering $a=0.2$, this state is separable for $\epsilon\leq0.56$.
In Figs. 3(a) and 3(b) we show the corresponding trajectories for
separable states, while in Figs. 4(a) and 4(b) we show the trajectories
for entangled states. Note that the same behavior as that for $a=b=1/\sqrt{2}$
can be observed in Fig. 5(b) that consider $a=0.2$: given the same
initial conditions, quantum trajectories for separable states are
smooth and oscillate less than those corresponding to entangled states.
For a complete mixture ($\epsilon=0$), the trajectories are straight
lines. 

Now let us see what happen to Bohmian trajectories in the presence
of losses. Again using Eq. (\ref{2-2-1}), we can check that considering
$a=b$ and $\epsilon=0.4$ this state is separable at $\gamma t=0.15$.
Figs. 6(a,b) and 7(a,b) show the corresponding trajectories for damped
states. As expected, the reservoir attenuates the oscillations, that
turn to be more accentuated when the damping rate is larger.

Note that these results for damped states show the same behavior as
that for undamped states: given the same initial conditions, quantum
trajectories for separable states are smooth and the amplitude of
oscillations is less than those corresponding to entangled states,
with the amplitude of oscillations going to zero when $\epsilon\rightarrow0.$

As a final remark, it is worth noting that Bohmian trajectories may
provide a way towards detecting quantum separability of mixed quantum
states. As a matter of fact, when the interpolator parameter $\epsilon$
evolves from $0$ to $1$, the trajectories changes from straight
lines to curves with steep slopes. However, around the regions where
the trajectories of the particles come closer together ---thus interfering
to a greater extent--- their curvatures become very smooth. By focusing
our attention on these regions of maximum interference between the
trajectories, which takes place in Figs. 2 and 4 for $\omega t=(2n+1)\pi/2$,
$n=0,1,2,...$, we thus observe that when the parameter $\epsilon$
evolves from $0$ to $1$, the slopes of the trajectories starts from
$0$, seems to reach a maximum value and then decreases due to the
strong interference between the particles paths. We might suspect
that the maximum curvature takes for the value of $\epsilon$ that
gives the separability condition for the density matrix. However,
a problem arise when we set out to compute the curvature of the trajectories
(in the specified regions) as a function of the parameter $\epsilon$,
since the curvatures of the trajectories are different for different
initial positions $\tilde{x}_{\alpha}$ of the particles.

\section{Conclusions}

In this paper we have derived Bohmian trajectories for noninteracting
bipartite states of damped harmonic oscillators under a thermal reservoir
at finite temperature in a similar way to that of Vink's extension
of Bell's beables\cite{Vink}. As an application, we have calculated
the trajectories for a generalized Werner state dissipating at zero
temperature in regions where the two systems are either entangled
or separable according to Wooters' concurrence. Our results indicated
that individual trajectories for entangled states differ slightly
in the amplitude of oscillation as compared with those corresponding
(same initial conditions) trajectories for disentangled states. We
note, however, that according to our simulations, this difference
is not enough to characterize unambiguously separability or entanglement,
which is a global property of the system. This is so because the trajectories
change continuously when the state changes from separable to nonseparable.
We hope these preliminary results can encourage future research towards
an eventual link between separability and Bohmian mechanics.

\begin{center}\textit{\textbf{ Acknowledgements } }\end{center}

The authors acknowledge the support from FAPESP, CNPQ, CAPES, and
INCT, Brazilian agencies. A. R. de Almeida thanks Fapeg for partial
support.

\end{document}